# Detector dead-time effects and paralyzability in high-speed quantum key distribution

**Daniel J Rogers[1, 3], Joshua C Bienfang[1], Anastase Nakassis[2], Hai Xu[2] and Charles W Clark[1]**

[1]Joint Quantum Institute,
National Institute of Standards and Technology and University of Maryland, Gaithersburg, Maryland 20899, USA

[2]Information Technology Laboratory
National Institute of Standards and Technology
Gaithersburg, MD 20899, USA

**Abstract:** Recent advances in quantum key distribution (QKD) have given rise to systems that operate at transmission periods significantly shorter than the dead times of their component single-photon detectors. As systems continue to increase in transmission rate, security concerns associated with detector dead times can limit the production rate of sifted bits. We present a model of high-speed QKD in this limit that identifies an optimum transmission rate for a system with given link loss and detector response characteristics.

## 1. Introduction

Recently there has been interest in developing single-photon quantum key distribution (QKD) systems that can support one-time-pad encryption at bit rates consistent with broadband telecommunications [1-6]. While the range over which secret cryptographic key can be produced by a QKD system is bounded by noise and losses in the quantum channel, below this bound it is generally true that an increase in the quantum-channel transmission rate results in an increase in the secret-key production rate [7]. For a given distance-bandwidth product, it is therefore possible to increase the range of continuous one-time-pad encryption services by increasing the quantum-channel transmission rate [8]. This approach has motivated the development of QKD systems operating at the highest transmission rate supported by better detector timing resolution. Improvements in detector timing jitter, particularly in silicon single-photon avalanche photodiodes (SPADs), have enabled the demonstration of systems with transmission rates above 1 gigahertz [9].

Silicon SPADs have a finite recovery time, $\tau$, that is typically of the order of 100 ns. This interval, known as the dead time, is the period after each detection event during which the device does not respond to another incident photon. The dead time occurs after a detection event triggers an avalanche in the SPAD. The amplified avalanche current must be quenched and free charge carriers must be removed from the SPAD before it can be reset to its active state. This process limits the maximum count rate of such devices to less than $\tau^{-1}$. It is worthwhile to note that while superconducting single-photon detectors (SSPDs) can support significantly higher count rates, they exhibit finite reset times due to kinetic inductance, albeit in the range of 1 ns to 10 ns [10]. For most QKD systems, dead-time effects are reasonably assumed to have a negligible impact on overall performance; typical transmission rates, $\rho_{TX}$, and link losses, $L$, are such that most systems operate in a regime where the detection rate is low with respect to the maximum count rate, i.e., $\rho_{RX} << \tau^{-1}$. As the rates of transmission and key production increase however, QKD systems will move out of this regime. In this article we present a model of QKD in the BB84 protocol [11] that describes both the count-rate limitations and the security issues that arise as key production rates increase with $\rho_{TX}$. We find that for given values of $L$ and $\tau$, there is a transmission rate that maximizes the sifted-bit rate. Contrary to naïve intuition, this maximum sifted-bit rate can be significantly greater than $(2\tau)^{-1}$.

In section 2, we describe how dead-time effects compromise secure QKD if transmission speeds are increased inattentively. Section 3 describes the secure operation of a high-speed QKD system, and presents an analytic model that characterizes the sifted-bit rate in terms of the link parameters. Section 4 gives specific results of this model, which are confirmed by Monte Carlo simulations. Section 5 describes additional hardware-based solutions to dead-time related issues.

## 2. Problems encountered in the high-speed regime

The most common detector configuration for QKD in the BB84 protocol is one in which the receiver, Bob, has a separate single-photon detector for each bit value in each basis. We restrict our discussion to this configuration and further assume that the detectors are free-running SPADs whose low noise allows them to be used without active gating. This is often the case in free-space QKD systems and fiber QKD systems with up-conversion detectors [1-6].

In this configuration, when the quantum-channel transmission rate satisfies $\rho_{TX} > \tau^{-1}$ photons can arrive and be detected at the receiver at a time when one or more of the SPADs is recovering from a prior detection event. If two such detection events occur in the same basis they necessarily correspond to opposite bit values and are completely correlated [12]. We address this obviously critical operational concern in the next section. For the purpose of demonstrating the influence of this effect, let us briefly consider an inappropriate implementation of high-speed QKD in which $\rho_{TX}$ is increased with complete disregard for the dead time, while $L$ and $\tau$ remain constant, and measure the correlations induced in the resulting bit string. Although it seriously underestimates the extent to which information is available to an eavesdropper, a common statistical measure of correlations in the sifted-bit string is the probability that two adjacent bits in the sifted-bit string will have different values. This transition probability, $P_{trans}$, is

$$P_{trans} = \frac{1}{N-1} \sum_{i=1}^{N-1} \left[ \left( \text{bit}[i] + \text{bit}[i+1] \right) \bmod 2 \right], \qquad (1)$$

where bit[$i$] is the $i^{th}$ bit in the sifted-bit string, $N$ is the total number of sifted bits and the addition is performed modulo 2. From the description above, as $\rho_{TX}$ increases, $P_{trans}$ will tend to increase from 0.5 for uncorrelated bits. These dead-time induced correlations are demonstrated in figure 1, which shows results from a Monte-Carlo simulation of traditional BB84 QKD without applying any techniques to mitigate the effects of the detector dead time. Clearly $P_{trans}$ increases from 0.5 as the number of transmission periods per dead time, $k \equiv \tau \rho_{TX}$, increases above 1, approaching a value of 0.622 for BB84 as configured above. Increasing the link loss, thereby reducing the count rate at the receiver, reduces the value of $P_{trans}$ but does not insulate the system from dead-time effects.

The asymptotic value of 0.622 is unique to BB84 with four detectors and can be understood from the following calculation [15]. At high photon-arrival rates, detection events tend to occur in fixed sequences; the detectors recover and then fire again in order. Without loss of generality, we arbitrarily choose one detector to produce the first sifted bit. After this event there are six possible detection sequences, which are listed in table 1. Consider as an example the detection sequence 1-3-4-2, with detectors 1 and 3 representing bit value '1' in their respective bases and detectors 2 and 4 representing '0' in their respective bases. For this ordering, the probability, $P_3$, that the next detection event on detector 3 will produce the next sifted bit is $P_3 = (\frac{1}{2})^1$. Similarly, $P_4 = (\frac{1}{2})^2$, $P_2 = (\frac{1}{2})^3$, $P_1 = (\frac{1}{2})^4$, and so on, are the probabilities that detectors 4, 2 and 1 respectively will produce the next sifted bit *after* detector 1 produces the first sifted bit (i.e., all detection events in between are *not* included in the sifted key). Given the sifted-bit value of '1' from the first detection event and the subsequent infinite sequence of the 1-3-4-2 firing order, we calculate the probability that next sifted bit is a '0' to be:

$$P_{1342}(0) = \sum_{n=0}^{\infty} \left[ \left( \frac{1}{2} \right)^2 + \left( \frac{1}{2} \right)^3 \right] \left( \frac{1}{2} \right)^{4n} = \frac{2}{5} \ , \qquad (2)$$

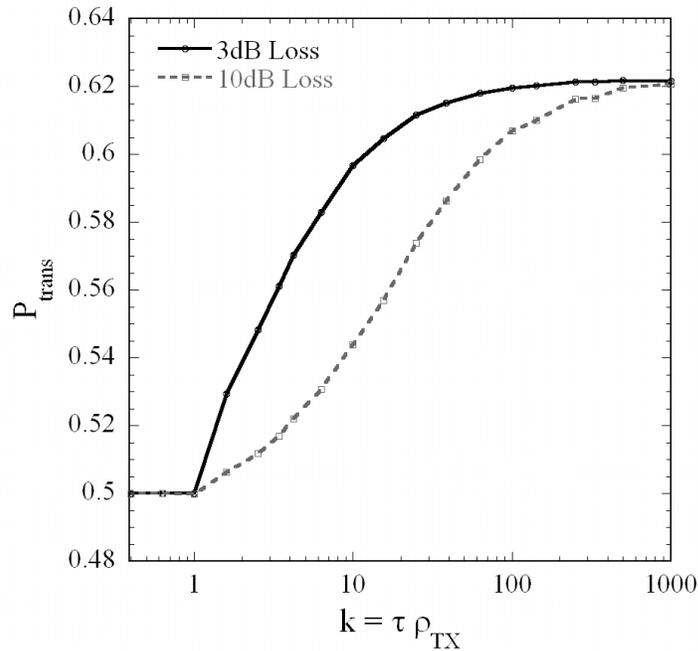

**Figure 1.** Bit-value transition probability vs. the number of transmission periods per dead time for two values of the link loss. The sifted bits used here were produced by Monte Carlo simulation and clearly demonstrate the onset of dead-time effects at transmission rates $\rho_{TX} > \tau^{-1}$.

**Table 1.** Individual transition probabilities for each detection sequence. Boldface indicates a detection event that corresponds to a '0' bit value.

| Detection Sequence | Transition Probability |
|---|---|
| 1-**2**-3-**4** | $\sum_{n=0}^{\infty}\left(\frac{1}{2}+\frac{1}{8}\right)\left(\frac{1}{2}\right)^{4n}=\frac{2}{3}$ |
| 1-**2**-**4**-3 | $\sum_{n=0}^{\infty}\left(\frac{1}{2}+\frac{1}{4}\right)\left(\frac{1}{2}\right)^{4n}=\frac{4}{5}$ |
| 1-3-**2**-**4** | $\sum_{n=0}^{\infty}\left(\frac{1}{4}+\frac{1}{8}\right)\left(\frac{1}{2}\right)^{4n}=\frac{2}{5}$ |
| 1-3-**4**-**2** | $\sum_{n=0}^{\infty}\left(\frac{1}{4}+\frac{1}{8}\right)\left(\frac{1}{2}\right)^{4n}=\frac{2}{5}$ |
| 1-**4**-**2**-3 | $\sum_{n=0}^{\infty}\left(\frac{1}{2}+\frac{1}{4}\right)\left(\frac{1}{2}\right)^{4n}=\frac{4}{5}$ |
| 1-**4**-3-**2** | $\sum_{n=0}^{\infty}\left(\frac{1}{2}+\frac{1}{8}\right)\left(\frac{1}{2}\right)^{4n}=\frac{2}{3}$ |

Repeating this calculation for all six possible sequences gives us the individual transition probabilities for each of the six detection sequences, shown in table 1. Since each sequence is equally likely to occur within a long sifted key, we average the resulting transition probabilities to obtain

$$P(0) = \frac{1}{6} \left( \frac{2}{3} + \frac{4}{5} + \frac{2}{5} + \frac{2}{5} + \frac{4}{5} + \frac{2}{3} \right) = 0.622 \ . \qquad (3)$$

### 3. Secure high-speed QKD

Since Eve, the nefarious eavesdropper, has access to the classical channel, she knows when bits are detected and in which basis they are sifted. As discussed above, when sequences of two or more detection events occur in a single basis with spacing less than the dead time, the detectors within a single basis fire alternately. This phenomenon provides Eve with nearly all of the information about the sifted bit string except for one bit representing which detector fired first within a given basis. Therefore such detection sequences, regardless of their length, can produce at most a single sifted bit. In other words, production of a sifted bit from a detection sequence of any length requires that the detection sequence begin when both detectors in a given basis are active. This requirement is necessary for the secure operation of a QKD system at transmission rates $\rho_{TX} > \tau^{-1}$ and must be imposed on the receiver either by some means of gating the detectors or by the sifting algorithm. We find the software solution to be both more practical and more efficient; we discuss active-gating and other hardware-based schemes in section 5.

In the low-count-rate regime the sifted-bit rate increases with the transmission rate. As the count rate at the receiver increases, the likelihood of closely spaced detection events in a single basis also increases. Continued increases in transmission rate eventually result in longer and longer detection sequences, each of which can result in at most a single sifted bit. This effect tends to reduce and eventually outweigh the potential improvement in sifted-bit rate gained by increasing the transmission rate. The model we present below shows that, for a given link loss and detector dead time, these competing effects create an optimum transmission rate above which further increasing the transmission rate results in a decrease in the sifted-bit rate.

We can calculate the probability that both detectors in a given basis are active when a photon is detected with the state-space model shown in figure 2. In this 2-dimensional model, the state of one of the receiver's bases, in this case the H-V basis, is quantified by how many clock periods need to pass before each detector in the basis is active, e.g., the state (3, 7) would denote that the H detector is three clock cycles away from being alive while the V detector is seven. Assuming that the two detectors have the same dead time, the state space ranges from 0 to $k = \tau \rho_{TX}$, as shown. On each transmission period, or clock cycle, a given detector either moves one period closer to recovery, or, if already active, the detector remains so or undergoes a detection event and moves $k$ periods away from recovery. The probability that both detectors are active is given by the probability $P_{0,0}$ that the basis is in the state $(0, 0)$.

To quantify the probability of having a detection event during a given clock cycle we find it useful to define a link loss parameter $p \equiv L/8$, where $L$ is the probability that a transmission event at Alice is detected at Bob (the detectors are assumed identical). The factor of 8 accounts for Bob's basis choice (1/2), and Alice's state choice (1/4). Therefore, ignoring noise, $p$ represents the probability that a particular detector produces a sifted bit on a given clock cycle. The probability that a particular detector fires on a given clock cycle is $2p$. As an example, we calculate the likelihood of the hypothetical detection sequence shown in figure 2. This sequence starts with a detection event on the "V" detector with probability $2p$, moving the basis from the origin to the state $(0, k)$. For the next four clock cycles the "H" detector does not fire with probability $(1-2p)^4$, followed by a detection event on the "H" detector with probability $2p$. The basis is now in the state $(k, 3)$ and both detectors are inactive. The state evolves with unity probability for three clock cycles until the "V" detector recovers, and then returns to the origin as the "V" detector does not fire for the next five clock cycles with likelihood $(1-2p)^5$. The probability of this particular hypothetical detection sequence is therefore $(2p)^2(1-2p)^9$.

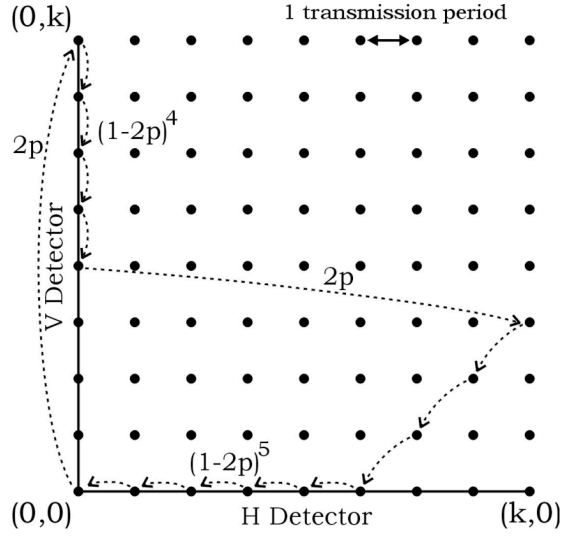

**Figure 2.** The state of Bob's H-V detection basis depicting a hypothetical detection sequence and associated probabilities. The size of the space is determined by the value of $k$, in this case chosen to be 8. Note that, although they are depicted, the diagonal states are not accessible in the absence of noise, since they can only result from the simultaneous detection of the same photon by both detectors.

The state-space picture allows us to calculate the probability $P_{0,0}$ that a given basis is in the state $(0, 0)$ as follows. We can write the probability that both detectors are active at the $(n+1)$ clock cycle as

$$P_{0,0}^{n+1} = (1-4p)P_{0,0}^n + (1-2p)P_{0,1}^n + (1-2p)P_{1,0}^n \,, \qquad (4)$$

where the first term represents the probability of no detection events occurring and the next two terms represent recovery from the (0, 1) and (1, 0) states respectively. We ignore recovery from the state (1, 1) because such diagonal states require simultaneous detection events that will not occur in the absence of noise. In steady state we drop the superscript and note that with random signals and identical detectors the steady-state behaviors of the H and V detectors are the same, allowing us to write $P_{0,1} = P_{1,0} = P_1$. Thus we find

$$P_1 = \left( \frac{2p}{1-2p} \right) P_{0,0} \ . \tag{5}$$

By the same argument one can write the probabilities $P_{k,0} = P_{0,k} = P_k$ as

$$P_k = (2p)P_{0,0} + (2p)P_1 \ , \tag{6}$$

which, with substitution for $P_{0,0}$ from (5), reveals that $P_k = P_1$. In fact, similar calculations for $P_{1,0}$, $P_{2,0}$, etc, show that all the states lying upon the axes, of which there are $2k$, have the same steady-state probability $P_1$.

The states not lying on one of the axes represent instances when both detectors are dead. Omitting the states along the diagonal, the internal states are only accessible from one of the on-axis states. Since the on-axis states are all of equal probability one can show that the internal states, of which there are $(k^2 - k)$, are also of equal probability, in this case $(2p)P_1$.

The expressions above represent the steady-state probabilities of the basis being in each of the states in the entire state space. We normalize the sum of these probabilities, giving

$$P_{0,0} + (2k)P_1 + \left( k^2 - k \right)(2p)P_1 = 1 \ . \tag{7}$$

Substituting for $P_1$ from (5) and solving for $P_{0,0}$ we obtain the steady-state probability that both detectors are alive for a given transmission event, as a function of $p$ and $k$:

$$P_{0,0}(p,k) = \left[ 1 + (2k)\left( \frac{2p}{1-2p} \right) + \left( k^2 - k \right)\left( \frac{(2p)^2}{1-2p} \right) \right]^{-1} \ . \tag{8}$$

As stated above, a detection sequence can only produce a sifted bit from events that occur when both detectors are alive. Therefore $P_{0,0}(p,k)$ should be used as an additional factor in the calculation of a system's sifted-bit rate. $P_{0,0}(p,k)$ is shown in figure 3 as a function of the normalized transmission rate $k$, for three values of the link loss $L = 8p$. It can be seen that as the transmission rate is increased $P_{0,0}(p,k)$ begins to roll off, approaching zero as $k^{-2}$ at high count rates. The roll-off of $P_{0,0}(p,k)$ marks the onset of dead-time effects and the departure from the low-count-rate regime.

The behavior of $P_{0,0}(p,k)$ in the high-count-rate regime illuminates a characteristic unique to operation of QKD systems at transmission rates $\rho_{TX} > \tau^{-1}$. From the standpoint of producing sifted bits, when the pair of SPADs in a given basis is considered as a whole, the QKD receiver becomes what is known as a paralyzable counter [13-15]. Signals that arrive at a paralyzable counter during recovery, though not counted, extend the necessary recovery time [20]. In contrast, non-paralyzable counting systems recover from each counting event regardless of signals that arrive during the dead time. Taken individually, SPADs are non-paralyzable detectors; with the exception of twilight counts, when the bias voltage is below the breakdown voltage, photons that arrive during the dead time have no significant effect on the detector [16]. It is worthwhile to note that the response of paralyzable and non-paralyzable systems exhibit significant differences only in the regime of high count rates [14], and paralyzability has become relevant to QKD systems as researchers seek to increase key-production rates.

Although each closely-spaced detection sequence can produce at most a single sifted bit, it is also true that as the length of the detection sequence grows the likelihood that a bit will be sifted from the sequence also grows. In the low-count-rate regime the average length of a detection sequence is 1 and the likelihood of sifting a bit from a sequence is 0.5. For a detection sequence of length 3, however, the likelihood that at least one of the detection events occurred in the correct basis is 7/8. This fact offsets some of the deleterious dead-time effects and must be included in the calculation of the sifted-bit rate.

At any count rate, we can write the probability of sifting a bit from a detection sequence that begins when both detectors were active can as

$$S(p,k) = \sum_{N=1}^{\infty} \left( 1 - \left( \frac{1}{2} \right)^N \right) T_N(p,k) \,, \tag{9}$$

where $T_N(p,k)$ is the probability that the detection sequence consists of $N$ detection events. To calculate $T_N(p,k)$ we can use the state-space model from figure 2. For a detection sequence of length 1 (i.e. a single detection event) there is only one path through the state space. For longer sequences we must sum the possible paths for a given number of detection events.

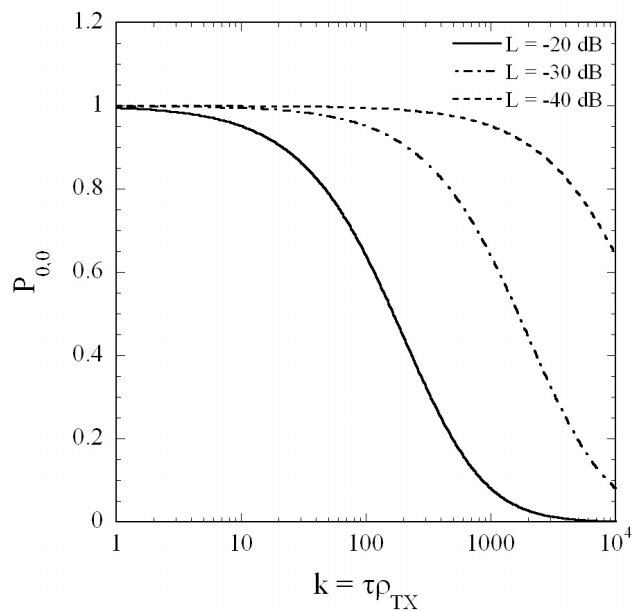

**Figure 3.** The likelihood $P_{0,0}$ that both detectors in a given basis are active when a photon arrives vs. the number of transmission periods per dead time, $k$, for three values of the link loss $L$. The fact that $P_{0,0}$ tends to zero at high transmission rates demonstrates the paralyzability of the QKD receiver.

For example, there are a total of *(k-1)* ways to arrange two detection events before the basis returns to the state (0, 0); only one of these ways is depicted in figure 2. The probabilities of a detection sequence having length up to $N = 4$ are

$$T_1(p,k) = (1 - 2p)^k \ , \tag{10}$$

$$T_2(p,k) = \sum_{\sigma=0}^{k-1} 2p(1 - 2p)^{2\sigma+1} \ , \tag{11}$$

$$T_3(p,k) = \sum_{\sigma_1=0}^{k-1} \sum_{\sigma_2=0}^{\sigma_1} (2p)^2 (1 - 2p)^{2\sigma_2+k} \ , \tag{12}$$

$$T_4(p,k) = \sum_{\sigma_1=0}^{k-1} \sum_{\sigma_2=0}^{\sigma_1} \sum_{\sigma_3=0}^{k-(\sigma_1-\sigma_3)-1} (2p)^3 (1 - 2p)^{2(\sigma_1+\sigma_3)+1} \ . \tag{13}$$

The truncated geometric series in $T_N(p,k)$ can be evaluated with standard techniques to yield analytic expressions for all $N$. While the sum over $N$ in (9) is theoretically infinite, it is worthwhile to note that in practice one needs to compute $T_N(p,k)$ only up to $N = 6$, as the probability of sifting a bit from a detection sequence longer than six events approaches unity.

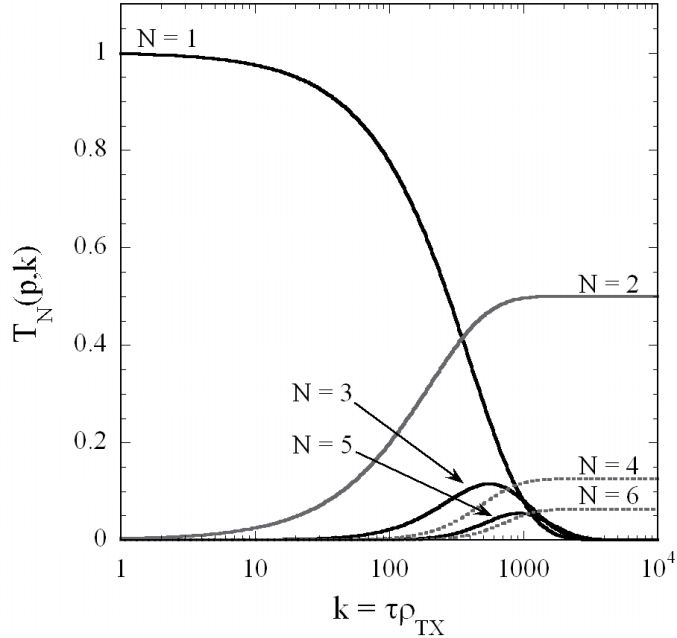

**Figure 4.** The probabilities $T_N(p,k)$ of a detection sequence having $N$ detection events vs. the normalized transmission rate $k$, for link losses $L = $ -20 dB. There is a characteristic difference in even and odd numbers of detection events in the high-count rate regime.

One interesting feature of $T_N(p,k)$ is the difference between even and odd values of $N$, as illustrated in figure 4. While all $N > 1$ sequences have low probability in the low count-rate regime, at high count rates the $N$:odd sequences fall asymptotically to zero, but the $N$:even sequences have constant finite probabilities. This behavior can be understood from the fact that an $N$:even sequence can minimize the number of clock cycles during which a detector is active but does not fire. For an $N$:odd sequence, the unlikely situation occurs where a live detector must not fire for at least $k$ clock cycles before the basis returns to $(0, 0)$.

Noise sources such as background counts and detector dark counts can also be included in the model in a straightforward manner. We define $\varepsilon$ as the probability that a detector experiences a noise event during one clock cycle. The probability that a detector fires during a clock cycle therefore changes from *(2p)* to *(2p + ε)*, which can be substituted into $P_{0,0}(p,k)$ and $T_N(p,k)$ accordingly. As mentioned above, a noise event on one detector can occur on the same clock cycle as a signal (or noise) event on the other detector. These simultaneous events put the basis in the state $(k, k)$, after which the basis recovers with unity probability along the diagonal back to $(0, 0)$. Thus we find that when noise counts are included, the probability that both detectors are alive on the $(n+1)$ clock cycle becomes

$$P_{0,0}^{n+1} = \left(1 - 2\left(2p + \varepsilon\right)\right)P_{0,0}^n + 2\left(1 - \left(2p + \varepsilon\right)\right)P_1^n + \left(2p\varepsilon + \varepsilon^2\right)P_{0,0}^{n-k} \ , \tag{14}$$

where the third term accounts for simultaneous detection events. The steady state calculation of $P_{0,0}(p,k)$ proceeds in the same manner as described above. It should be noted that while noise can cause simultaneous detection events, no secure bits can be sifted from such events.

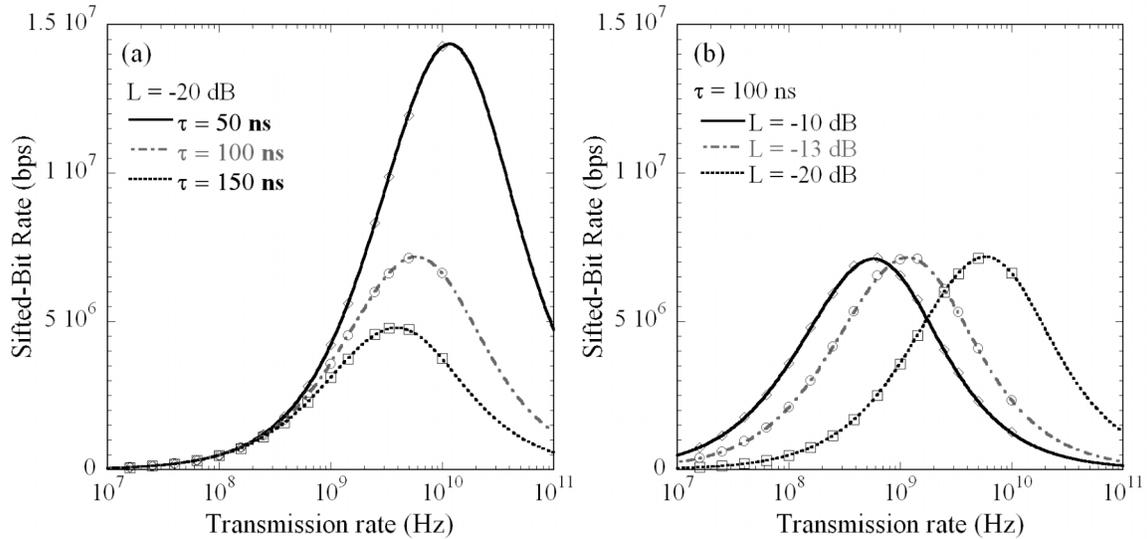

**Figure 5.** The sifted-bit rate including dead-time effects, showing excellent agreement between the model (lines) and the simulation (symbols). The effect of varying the dead time with fixed link loss is shown in (a). The effect of varying the link loss with fixed dead time is shown in (b).

## 4. The sifted-bit rate in high-speed QKD

We have described all the factors necessary to incorporate dead-time effects in to the sifted-bit production rate. Returning to the noiseless picture, we write the sifted-bit rate as

$$SBR = \rho_{TX} \, 8p P_{0,0}(p,k) S(p,k) \,, \tag{15}$$

where $\rho_{TX}$ is the transmission rate, $8p$ is the link loss, $P_{0,0}(p,k)$ is the probability that both detectors are active, given by (8), and $S(p,k)$ is the likelihood of sifting a bit from a detection sequence, given by (9). The sifted-bit rate is shown in figure 5 as a function of the transmission rate for various detector dead times (a) and link losses (b). The lines indicate the results from the analytic state-space model presented above. The symbols indicate results from a BB84 Monte-Carlo simulation that incorporates the modified sifting algorithm described above – sifting at most a single bit from sequences of closely-spaced detection events. The simulation generated 1 MB of sifted bits at each point using an ANSI C standard random-number generator running on a Linux computing cluster. As illustrated in figure 5, dead-time effects induce a maximum value on the sifted-bit rate, above which further increases in transmission rate reduce the sifted-bit rate. The maximum value of the sifted-bit rate is a complicated function of the link parameters. However, we find this maximum is not strongly dependant on the link losses. As demonstrated in figure 5(b), it may be accurately approximated as a function of dead time alone by

$$SBR\big|_{\text{Max}} \approx \frac{1.433}{(2\tau)} \,, \tag{16}$$

where the constant of proportionality was found by a least-squares fit. The factor $(2\tau)^{-1}$ represents the maximum sifted-bit rate for the case of an actively gated receiver in which all the detectors are disabled when any one of them fires [12].

The transmission rate at which the sifted-bit rate is maximized is also a complicated function of the dead time and link losses. However, for typical link losses and detector dead times we find that it may be approximated to an accuracy better than 1% by

$$\rho_{TX}^{\max} \approx \frac{5.92}{8p\tau} \,. \tag{17}$$

For most QKD links the loss and dead time are such that detector timing resolution plays a dominant role in determining the optimum transmission rate [1, 2, 3, 5]. However, as the disparity between detector timing resolution and recovery time grows with improved timing resolution, transmission-rate limitations imposed by dead-time effects will become more significant.

## 5. Hardware approaches to addressing dead-time effects

There are a variety of methods that may be employed to address the security issue that arises in the $\rho_{TX} > \tau^{-1}$ regime. The algorithmic solution modeled above is, to our knowledge, the most efficient with respect to the production of sifted bits. An active hold-off scheme has been previously proposed, in which all the detectors are disabled when any one of them fires [12]. Actively disabling the detectors by some electronic means can be technically challenging, particularly as transmission rates exceed 1 GHz. As an alternative we propose the self-disabling receiver shown in figure 6. In this system the states in each of Bob's bases are sent to the same detector, though with different propagation delays depending on the state. The states of the photons incident on each detector are distinguished by their arrival times, much in the same manner that time-division-multiplexed (TDM) communications links distinguish various channels. With only one detector in each basis, the entire basis is disabled for the duration of the dead time, and sequences of closely-spaced detection events are eliminated. This QKD receiver is a non-paralyzable counter capable of producing sifted bits at rates up to $\tau^{-1}$ in the high-count-rate regime.

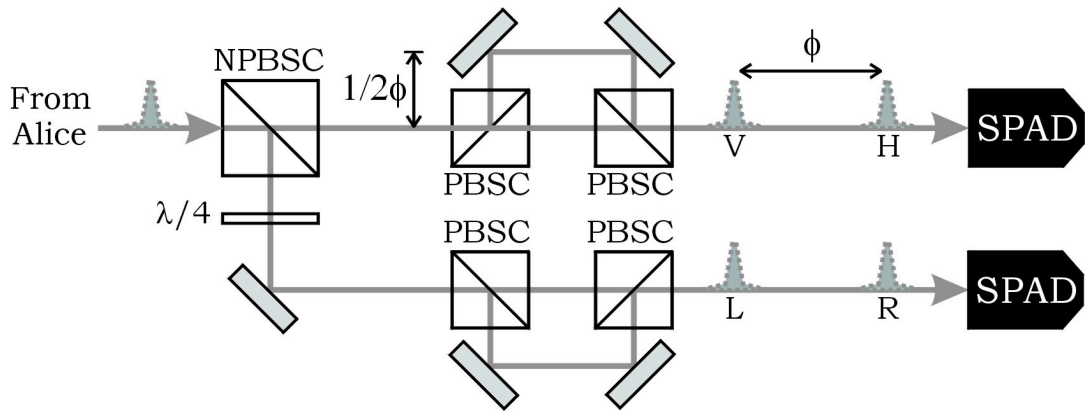

**Figure 6.** A BB84 receiver with self-disabling bases. In this configuration the individual states in each measurement basis are distinguished by their arrival times at the SPADs. (N)PBSC is a (non) polarizing beam-splitting cube.

As a consequence of operating the receiver bases in the self-disabling format shown in figure 6, each transmission event from Alice is analyzed in two time bins at Bob's receiver. If these time bins are limited by the SPAD's ability to distinguish photon-arrival times, that is, by the SPAD timing resolution, then Alice's transmission period must be at least twice as long. Thus the maximum transmission rate as determined by the detector timing resolution is reduced by one half. The reduction in transmission rate makes this type of receiver useful only in cases in which the algorithmic implementation described above is somehow impractical.

## 6. Conclusions

We have presented a model for the sifted-bit production rate of BB84-type QKD systems operating at transmission rates $\rho_{TX}$ that are higher than the maximum count rate of the component single-photon detectors. This model addresses critical security concerns that must be considered when operating in this regime and quantifies the onset of dead-time effects. We have established that, with free-running SPADs, high-speed QKD systems are paralyzable counting systems. This phenomenon emerges from the collective behavior of the pair of detectors in a given basis, as SPADs are non-paralyzable counting systems when considered individually. We have shown with both analytic modeling and Monte-Carlo simulation that dead-time effects cause there to be an optimum transmission rate that maximizes the sifted-bit production rate. The functional dependence of the maximum sifted-bit rate on the link parameters has been presented, and these relations will be useful in the design of QKD systems and single-photon detection systems.

This article has focused on polarization-encoded BB84 QKD. A useful extension of the analysis presented here would be the application of the state-space model to other protocols and encoding schemes. In particular, the differential-phase-shift encoding scheme used in Ref. [3] readily lends itself to extremely high transmission rates. Detector dead times are likely to have significant influence on the performance of such systems, and the analysis of such influence in the context of the current understanding would be useful.